\documentclass[12pt,preprint]{aastex}

\lefthead{Bekki, Couch, \& Shioya}
\righthead{Formation of nucleated dwarfs}

\begin{document}
\title{Dissipative transformation of non-nucleated dwarf galaxies
into nucleated systems}

\author{Kenji Bekki,    Warrick J. Couch} 
\affil{
School of Physics, University of New South Wales, Sydney 2052, Australia}

\and

\author{Yasuhiro Shioya}
\affil{Astronomical Institute, Tohoku University, Sendai, 980-8578, Japan}

\begin{abstract}

Recent photometric observations by the {\it Hubble Space Telescope (HST)} 
have revealed the physical properties of stellar galactic
nuclei in nucleated dwarf galaxies in the Virgo cluster of galaxies. 
In order to elucidate the formation processes of nucleated dwarfs,
we numerically investigate gas dynamics, star formation, 
and chemical evolution 
within the central 1 kpc  of gas disks embedded within the
galactic stellar components of non-nucleated dwarfs. 
We find that high density, compact stellar systems
can be formed in the central regions
of dwarfs as a result of  dissipative, repeated  merging of massive stellar
and gaseous clumps 
developed  from  nuclear gaseous spiral
arms as a result of local gravitational instability.
The central stellar components are found to have 
stellar masses which are typically $~$5\% of their host dwarfs 
and show very flattened shapes, rotational kinematics,
and central velocity dispersions significantly 
smaller than those of their host dwarfs.
We also find that more massive dwarfs can develop more massive,
more metal-rich, and higher density
stellar systems in their central regions,
because star formation and chemical enrichment proceed
more efficiently owing to the less
dramatic  suppression of star formation by  supernovae feedback effects
in more massive dwarfs.
Based on these results, we suggest that gas-rich, non-nucleated dwarfs
can be transformed into
nucleated ones  as a result of dissipative gas 
dynamics in their central regions.
We discuss the origin of the observed correlations between
physical properties of stellar galactic nuclei and
those of their host galaxies.

\end{abstract}


\section{Introduction}

A growing number of photometric studies of dwarf galaxies 
by the {\it Hubble Space Telescope (HST)} 
have recently  revealed the physical properties of stellar galactic nuclei
of dwarfs in the Virgo cluster
of galaxies (e.g.,  De Propris et al. 2005; Grant et al. 2005;
Ha\c segan et al. 2005). 
{\it HST}  photometric observations
have also revealed that a significant fraction of late-type galaxies
have very compact stellar systems
in the very center of these
galaxies (Phillips et al. 1996; Carollo et al. 1998; Matthews et al. 1999;
B$\ddot{\rm o}$ker et al. 2004; Walcher et al. 2005).
The physical properties of stellar galactic nuclei (e.g. masses and surface brightness) have been suggested to correlate with
those of their  host galaxies
(B$\ddot{\rm o}$ker et al. 2004).
Spectrophotometric studies of nucleated dwarf galaxies
have revealed the structural and  kinematical properties of 
their  central stellar components   
(e.g., Stiavelli et al. 2001; Geha et al. 2002; Lotz et al 2004).

Although these recent observations, combined with previous ones on 
nucleated dwarfs
(e.g., Binggeli \& Cameron 1991;
Ferguson \& Binggeli 1994),
have provided vital clues to understanding the formation
of nucleated dwarf ellipticals (dE,Ns)
and  irregulars (dI,Ns), only a small number of theoretical works have so far investigated the formation processes of their stellar galactic nuclei
(e.g., Tremaine et al. 1975;
Oh \& Lin 2000; Bekki et al. 2004).
One of the most extensively discussed scenarios is
that these stellar nuclei are formed from the merging of smaller stellar clusters of globular cluster-like mass in the central regions of galaxies
(e.g., Oh \& Lin 2000; Bekki et al. 2004).
These numerical works, which are based on {\it dissipationless} models
of merging star clusters, have provided some predictions on the structural and
kinematical properties of stellar nuclei.
These previous works, however,   
did not discuss at all how gaseous dissipation and
star formation are important in the formation of nucleated dwarfs, 
even though a number of spectrophotometric observations have already
provided possible evidence for very young stellar populations
in the stellar nuclei of some nucleated galaxies 
(e.g., Caldwell \& Bothun 1987).

The purpose of  this {\it Letter} is thus to
demonstrate, for the first time, that
gaseous dissipation and star formation
can play a vital role 
in the formation of nucleated dwarfs (i.e., dE,Ns and dI,Ns).
Based on the results of 
chemodynamical simulations of the gas dynamics and star
formation within the central 1kpc  of non-nucleated dwarfs,
we demonstrate that non-nucleated dwarfs
can be transformed into nucleated ones as a result of
{\it dissipative nucleation}.
We investigate the
structural, kinematical, and chemical properties of 
stellar galactic nuclei in nucleated dwarfs
with different masses and gas mass fractions and thereby provide
theoretical predictions that can be tested against recent
observations.
Although the present study suggests the importance of
gas clump formation in inward mass-transfer processes
for dwarfs,
this importance has already been proposed by Shlosman \& Noguchi (1993)
and Noguchi (1999) {\it in the context of triggering mechanisms of 
starbursts and bulge formation.}

\section{The model}

We  investigate the dynamical evolution of 
stellar and gaseous components within 
the central 1 kpc of a galaxy, 
via  chemodynamical simulations carried out
on a GRAPE board (Sugimoto et al. 1990). 
Since the numerical methods and techniques we use 
for the GRAPE system
have already been described in detail
elsewhere (Bekki \& Shioya 1999),
we give only a brief review here.

Old stellar components  of dwarf galaxies 
(dIs and dEs)
are observed to have exponential
luminosity profiles 
(e.g., Ichikawa, Wakamatsu, \& Okamura 1986)
and therefore the projected radial density profile  of
the stellar component  with mass $M_{\rm sph}$ 
and scale length $a_{\rm sph}$ is
assumed to be an exponential one with  the central
surface density of ${\Sigma}_{\rm sph,0}$.
Guided by recent observations on the scaling relation
between ${\Sigma}_{\rm sph,0}$ and
$M_{\rm sph}$ 
(i.e., ${\Sigma}_{\rm sph,0} \propto {M_{\rm sph}}^{0.54}$;
Kauffmann et al 2003),  
we adopt the  $M_{\rm sph}-a_{\rm sph}$ relation
of $a_{\rm sph} = C_0 \times {M_{\rm sph}}^{0.23}$,
where $C_0$ is chosen such that the mass and the scale length 
of the Galactic exponential bulge can be reproduced reasonably
well. For example, this equation gives $a_{\rm sph}=200$pc
for $M_{\rm sph}=10^9 {\rm M}_{\odot}$.

The gas disk in a dwarf is modeled 
as a collection of discrete gas clouds
with  the observed mass-size
relationship of GMCs (Larson 1981).
The thin gas disk composed of these clouds 
is assumed to have an exponential profile
with the scale length being the same as $a_{\rm sph}$. The mass ratio
of the gas disk to the initial 
stellar component is regarded as a free parameter
($f_{\rm g}$) that ranges from 0.02 to 0.5.
Every pair of two overlapping gas clouds
is made to collide with the same restitution coefficient (Hausman \& Roberts 1984)
and lose its random kinematic energy through gaseous dissipation.

Field star formation
is modeled by converting  the collisional
gas particles
into  collisionless new stellar particles according to the algorithm
of star formation  described below.
We adopt the Schmidt law (Schmidt 1959)
with exponent $\gamma$ = 1.5 (1.0  $ < $  $\gamma$
$ < $ 2.0, Kennicutt 1998) as the controlling
parameter of the rate of star formation.
These stars formed from gas are called ``new stars'' (or ``young stars'')
whereas stars initially within a spheroid  are called ``old stars''
throughout this paper.
Chemical enrichment through star formation and supernovae feedback
during the stellar nucleus  formation  and  evolution 
is assumed to proceed  locally and the values of the return parameter
and the chemical yield are  
set to be 0.3 and 0.005, respectively.
About 10\% of the total energy of one type-II supernovae 
($\sim 10^{51}$ erg) is assumed to be converted into kinematic
energy of gas around the supernovae (Thornton et al. 1998)
for the Salpeter IMF. The feedback effects are stronger in
low-mass dwarfs for the adopted mass-size relation
in the present study.

We mainly investigate projected radial density
profiles (${\mu}_{\rm s}$)  of new and old stars 
in the central regions of spheroids. In order to compare 
the simulated profiles (${\mu}_{\rm s}$) with the observed $B-$band
surface brightness profiles  (${\mu}_{\rm B}$),
we use the  formula  of
${\mu}_{\rm B} = 27.05 -2.5 \log (I_{\rm B})$,
where $I_{\rm B}$ is the $B-$band surface brightness of a stellar
system measured in units of ${\rm L}_{\odot}$ pc$^{-2}$.
We estimate $I_{\rm B}$ from ${\mu}_{\rm s}$ for a given mass-to-light-ratio
($M/L_B$) in the $B-$ band.
We assume that $M/L_B$ for old stars 
[${(M/L_B)}_{\rm old}$] is 10, corresponding to
a single stellar population
(SSP) with age 10\,Gyr and solar metallicity 
(Vazdekis et al. 1996),
and that for new stars, 
${(M/L_B)}_{\rm new}$ is 0.5, corresponding to an SSP with age 0.5\,Gyr
and solar metallicity.

We firstly describe the results of the
fiducial model with $M_{\rm sph}=10^9 {\rm M}_{\odot}$
and $f_{\rm g}=0.2$, which shows typical behaviors of dissipative
nucleus formation in dwarfs.
Then we describe briefly the dependences of the results on
$M_{\rm sph}$ and $f_{\rm g}$.
The details of the parameter dependences (including those on
$M/L_B$) will be given in our forthcoming papers. 
The total number of particles is $10^5$ for the old stars ($N_{\rm old}$)
and $2\times 10^4$ for the initial gas ($N_{\rm gas}$),  
and the time integration of
the equation of motion is performed by using the 2nd-order leap-frog method.

\section{Result}


Figure 1 shows how a compact stellar system is 
formed in the central region of the star-forming gas disk
embedded in the old stellar spheroid for the fiducial model.
Numerous spiral arms composed only of gas clouds first develop
owing to the stronger self-gravity of gas ($f_{\rm g}=0.2$)  as the gas disk
rotates ($T=62$ Myr).
Because of the enhanced cloud-cloud collision rate and the resultant efficient
gaseous dissipation within the gaseous spiral arms,
small gas clumps composed of several tens of gas clouds
can be formed within each spiral arm ($T=250$ Myr).
Some of these small gaseous clumps
are massive, compact, and strongly self-gravitating so that
gas can be continuously converted into stars within the clumps ($T=250$ Myr).
These clumps can spiral in owing to dynamical friction against
the stellar background of old stars (i.e., the stellar spheroid),
and consequently merge with one another to form two very massive clumps
composed mostly of new stars ($T=250$ Myr).

The two massive clumps that develop within the central 200\,pc
can interact with each other to form  tidal arms composed only
of new stars ($T=374$ Myr).
Their mutual tidal interaction results in the loss of orbital
angular momentum, 
and they therefore finally merge with each other to form a single compact stellar system
in the central region of the stellar spheroid.
Although the single nucleus is initially not 
in the very center of the spheroid,
dynamical friction between the nucleus and the stellar background
can finally transfer the nucleus into the mass  center of the spheroid
within $\sim 100$ Myr.
Thus a stellar galactic nucleus composed of new stars can be formed
from merging between massive stellar and gaseous clumps
in the central region of the spheroid.

Figure 2 clearly shows that the non-nucleated dwarf
is transformed into a nucleated system   
with a very high central surface brightness 
owing to dissipative formation of the compact stellar
system in its center.
The derived large differences 
in ${\mu}_{B}$ between old and new stars 
results from the combined effect of the central high
stellar density (due to dissipative stellar nucleus formation)
and the small $M/L$ of the young stellar population
in the nucleus.
Figure 3 shows the radial profiles of
the velocity dispersion ($\sigma$) estimated separately for the old
stars and the new stars at $T=686$Myr.
It is remarkable that the new stars show a velocity dispersion profile that increases radially from a central value of ${\sigma}_{0}\sim 15$ km s$^{-1}$, 
whereas the old stars show a nearly constant profile
with ${\sigma}_{0}$ of $\sim 60$ km s$^{-1}$.
The essential reason for the lower ${\sigma}_{0}$  
in the new
stars is that the stellar nucleus, which is formed from {\it dissipative
merging of clumps},  can be strongly  self-gravitating
so that  ${\sigma}_{0}$ is determined by the nucleus itself
(i.e., by its smaller mass of $\sim 0.05$ in our simulation units).
The stellar nucleus shows a significant amount of rotation
($20-50$ km s$^{-1}$ within the central 100pc) and a very
flattened intrinsic shape.

Figure 4 shows the age-metallicity relation of the new stars that
are located within the central 50 pc (thus within the nucleus) of the
galaxy at $T=686$Myr.
The mean age and metallicity of this stellar nucleus are estimated
to be $3.7 \times 10^8 (=10^{8.57}$) yr and $-0.29$ in [Fe/H],
respectively.
The final metallicity of $-0.29$ is significantly higher
than the initial value ([Fe/H]$_{\rm i}$ = $-0.44$)
for the chemical yield of 0.005 and [Fe/H]$_{\rm i}$ = $-0.44$,
which means that the stellar nucleus is more metal-rich
and younger than the background stellar (spheroidal) components.
Figure 4 clearly indicates that the stellar nucleus
is composed of stars with different ages and metallicity:
The ages of the new stars ranges from
to $6.3 \times 10^6 (=10^{6.8})$ yr to
$6.8 \times 10^8 (=10^{8.8})$ yr
whereas the  metallicities
range from $-0.44$ to $-0.077$ in [Fe/H].
This wider range of ages and metallicities reflect the fact that
the stellar nucleus can form from merging between different clumps,
which have different star formation and chemical evolution histories.
It is clear from Figure 4 that the more metal-rich stars are
more likely to be younger.


The models with different $M_{\rm sph}$ 
($2.5 \times 10^7 {\rm M}_{\odot}$, 
$1.0 \times 10^8 {\rm M}_{\odot}$, 
$5.0 \times 10^8 {\rm M}_{\odot}$, 
and $1.0 \times 10^9 {\rm M}_{\odot}$)
and different $f_{\rm g}$ (=0.02, 0.05, 0.2, and 0.5)
are investigated in the present study.
The dependences of the physical properties of the compact stellar systems
(which can be regarded as stellar nuclei)
on $M_{\rm sph}$ and $f_{\rm g}$
are briefly summarized as follows:
Firstly, the total masses of the stellar nuclei  ($M_{\rm nuc}$)
in units of $ {\rm M}_{\odot}$
within 50 pc (5\% of an initial gas disk size)
is correlated with the
initial  stellar masses ($M_{\rm sph}$), 
and this correlation can be described as
 $ M_{\rm nuc} = 0.046 \times M_{\rm sph} -0.0030 $.
Here we apply a least-squares fit to the results of the models
that have different
$M_{\rm sph}$ and clearly show
stellar nucleus formation. If we make a least-squares fit to the ${\log}_{10} M_{\rm nuc} -
{\log}_{10} M_{\rm sph}$ relation, we can derive
the relation of
${\log}_{10} M_{\rm nuc} = 1.1 \times {\log}_{10} M_{\rm sph} - 2.5.$
Secondly, the mass fraction  
of the stellar nucleus ($f_{\rm nuc}$) in nucleated dwarfs
is correlated with $M_{\rm sph}$ 
and the correlation can be described as
$ f_{\rm nuc} = 0.021 \times 
(\frac{M_{\rm sph}}{10^9{\rm M}_{\odot}}) +0.023. $
Thirdly, the metallicities of stellar nuclei
are higher than those of their host galaxies in most models.
Accordingly stellar nuclei can be significantly redder than
their host galaxies after significant aging of their stellar
populations. More massive dwarfs can have more metal-rich stellar
nuclei.

Fourthly, stellar nuclei do not develop within a few Gyr in the models
with smaller $f_{\rm g}$ ($<0.05$), which
strongly suggests that self-gravity of gas is quite an important
factor for dissipative formation of stellar nuclei.
The central surface brightness (${\mu}_{0}$) of stellar nuclei is 
higher in models with higher $f_{\rm g}$, which implies
that the observed diversity in 
${\mu}_{0}$ between different dwarfs 
(e.g., De Propris et al. 2005) can be understood in 
terms of a different amount of gas within the dwarfs.
Fifthly, dense stellar nuclei are more likely to develop
in dwarfs with higher stellar densities owing to their deeper
gravitational potentials (i.e., weaker supernovae feedback effects).

\section{Discussions and conclusions}

It has long been known that more luminous dwarf galaxies (dE) are
more likely to contain stellar nuclei 
(e.g., van den Bergh 1986):
The number fraction of dE,Ns among dEs 
is almost 100 \% for $M_{\rm B} \approx -18$mag and about 20 \%
for $M_{\rm B} \approx -12$mag.
The present simulations have demonstrated that 
less luminous galaxies can have less remarkable stellar nuclei and some 
stellar nuclei 
can hardly be identified observationally as distinct nuclei.
They also have suggested that the origin of this dependence is closely
associated with the fact that stellar nucleus
formation is more strongly suppressed
by stronger feedback effects in less luminous galaxies.
These simulation results suggest that the origin of the observed
dependence of dE,N fraction on the host luminosities can be
understood in terms of the luminosity (or mass) dependence of
the effectiveness of supernovae feedback in stellar nucleus formation.

Recently Lotz et al. (2004) have revealed that some stellar nuclei of dE,Ns have
colors that are significantly bluer than those of their host galaxies
and can be due to recent star formation episodes.
Several previous spectroscopic observations
of stellar nuclei in dE,Ns also revealed possible evidences for younger stellar populations these stellar nuclei (e.g., Caldwell \& Bothun 1987; Bothun \& Mould 1988).
The present study suggests that if these apparently young stellar nuclei were formed dissipatively from gas, as our simulations have demonstrated,
they should be significantly more metal-rich than their host galaxies.
It is thus doubtlessly worthwhile for future spectroscopic observation
to investigate the metallicity differences 
between stellar nuclei and their host galaxies.

We have demonstrated that gas dynamics and star formation 
in the central 1 kpc of dwarfs can be important for the transformation
of non-nucleated dwarfs into nucleated systems.
One of advantages of the dissipative nucleus formation  scenario  
is that it can naturally explain how  massive
stellar and gaseous clumps,
 which are ``building blocks'' of stellar nuclei,
can be formed in the central regions of gas-rich dwarfs.
This scenario, however, has not yet provided clear explanations
on some of the key observed correlations between dynamical
properties of stellar nuclei and those of their host galaxies 
(e.g., B$\ddot{\rm o}$ker et al. 2004).

\acknowledgments
We are  grateful to the anonymous referee for valuable comments,
which contribute to improve the present paper.
K.B. and W. J. C. acknowledge the financial support of the Australian Research
Council throughout the course of this work.
The numerical simulations reported here were carried out on GRAPE
systems kindly made available by the Astronomical Data Analysis
Center (ADAC) at National Astronomical Observatory of Japan (NAOJ).

\clearpage


\newpage
\includegraphics[angle=-0.0,scale=0.9]{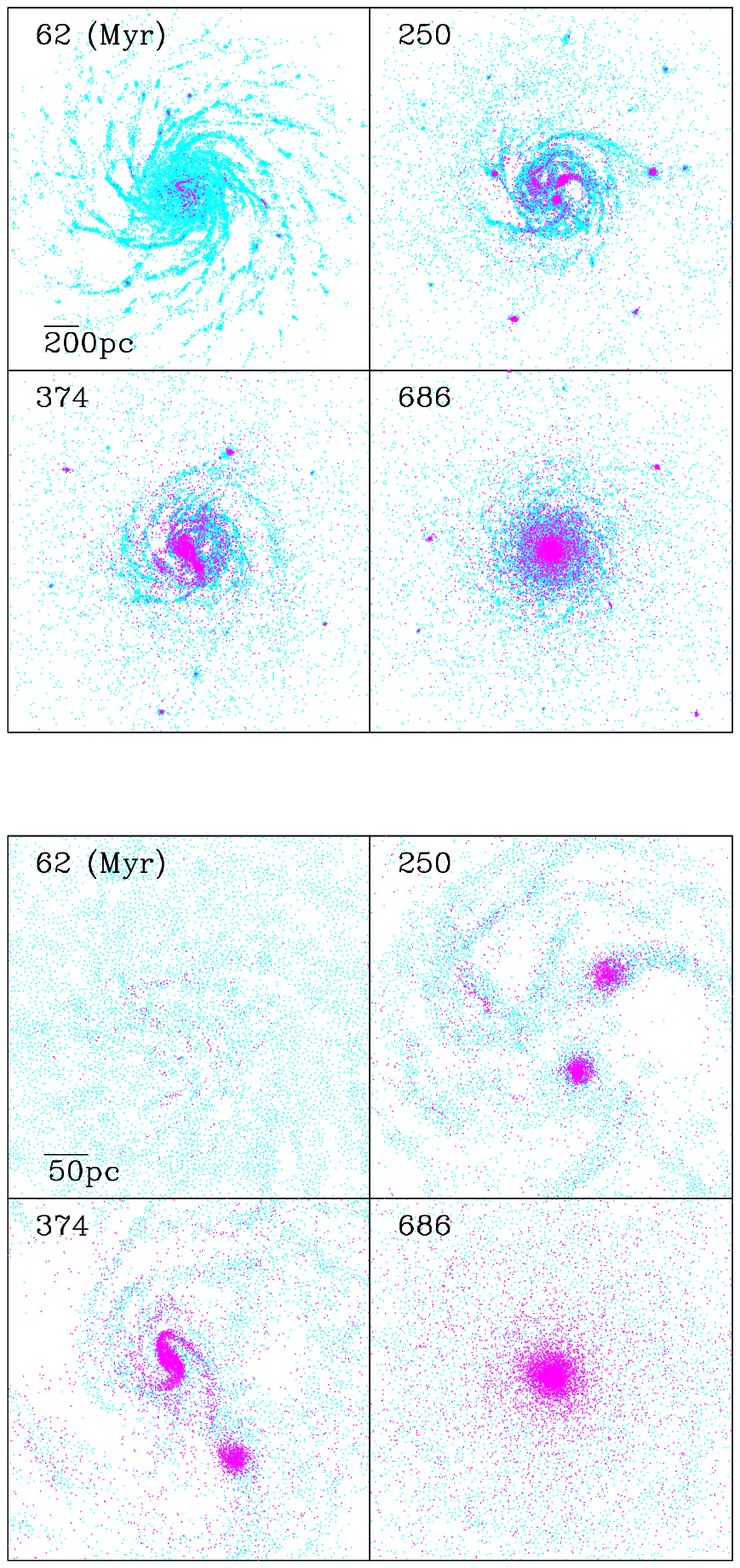}
\figcaption{
Morphological evolution of gas (cyan) and new stars (magenta)
in the fiducial model  projected onto the  $x$-$y$ plane.
The time $T$ shown in the upper left corner of
each panel  represents the time that has elapsed since
the simulation starts. One frame measures 2 kpc for the upper
four panels and  400 pc for the lower ones.
\label{fig-1}}

\newpage
\plotone{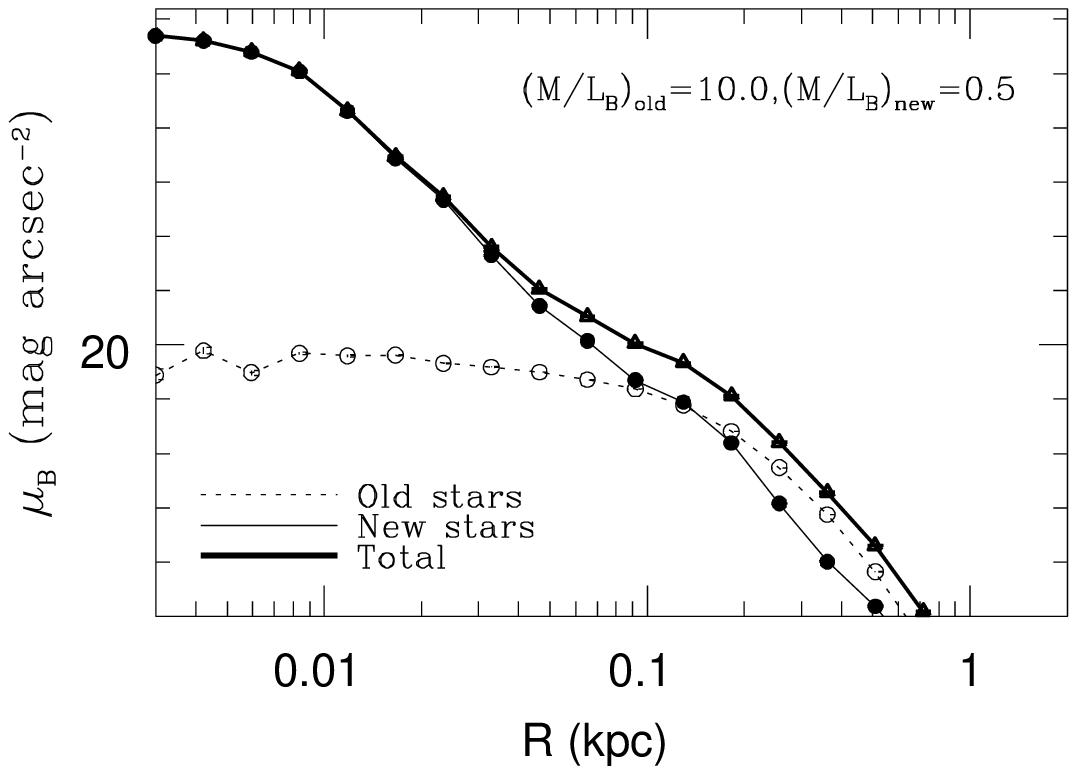}
\figcaption{
$B$-band surface brightness profiles (${\mu}_{\rm B}$)
for old stars (dotted), new stars (thin solid)
and total (i.e., old + new stars, thick solid) at $T=686$Myr
in the fiducial model with ${(M/L_{\rm B})}_{\rm old}=10$
and  ${(M/L_{\rm B})}_{\rm new}=0.5$.
The results correspond to the epoch
when the simulated nucleated dwarf is young ($\sim 0.5$ Gyr).  
\label{fig-2}}

\newpage
\plotone{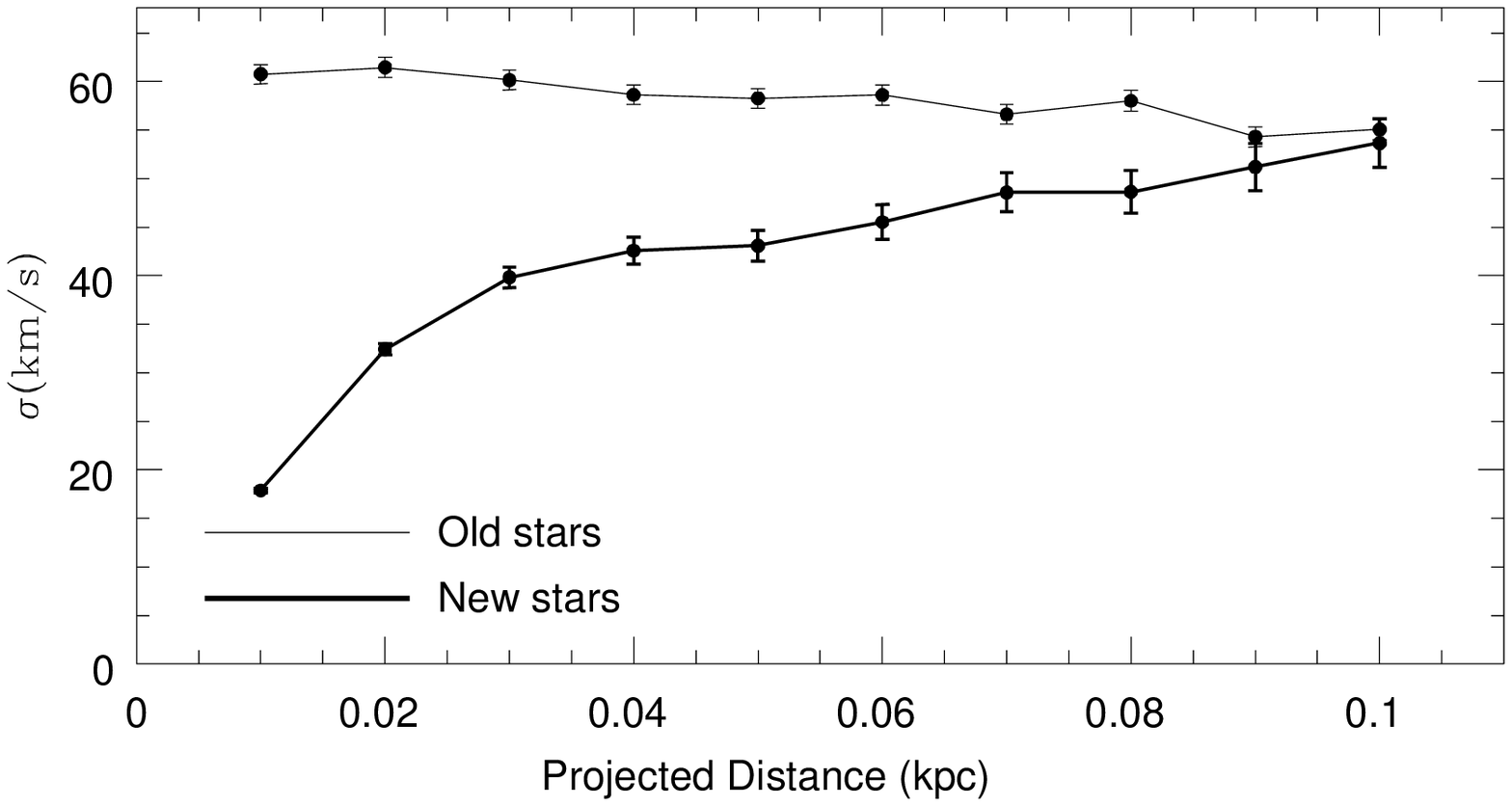}
\figcaption{
Radial profiles of velocity dispersion
for old stars (thin solid) and new ones (thick solid) at
$T=686$Myr for the edge-on view (i.e., projected onto the $x$-$z$ plane)
in the fiducial model.
\label{fig-3}}

\newpage
\plotone{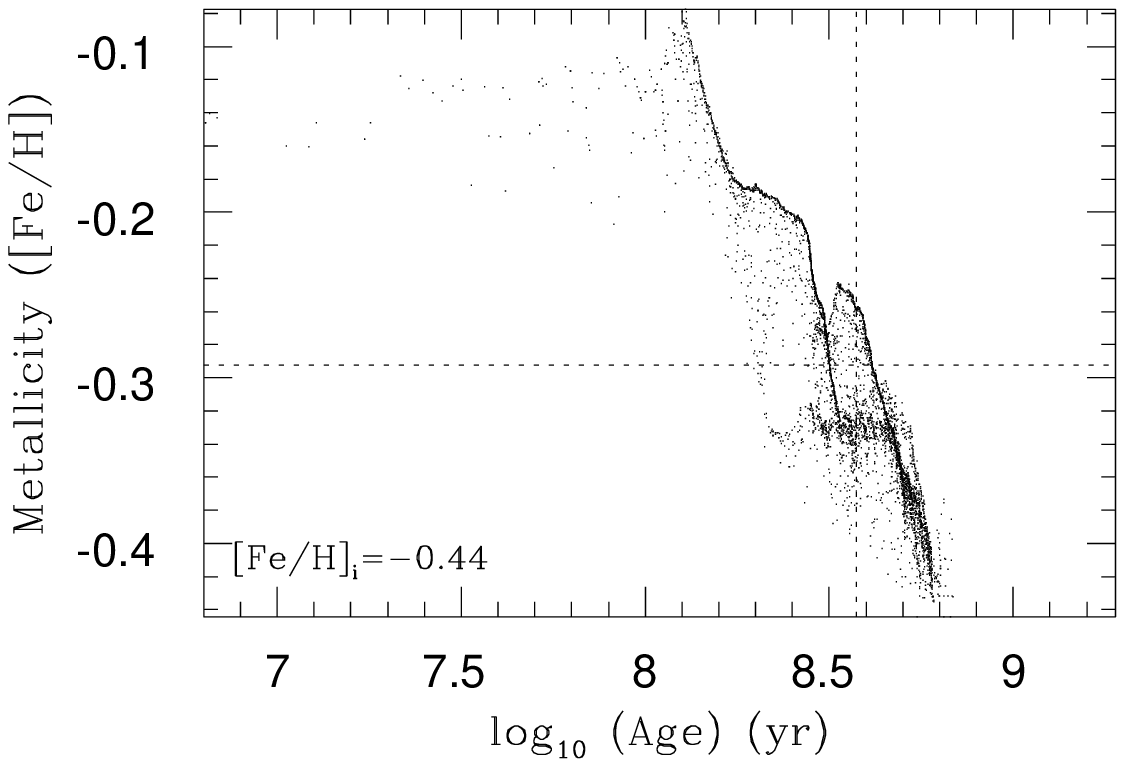}
\figcaption{
Distribution of new stars on the age-metallicity map 
at $T=686$Myr 
in the fiducial model  with the initial gaseous
metallicity ${\rm [Fe/H]}_{\rm i}$ of $-0.44$.
Here only new stars within the central 50 pc (thus those regarded
as being in the stellar nucleus) are plotted.
The horizontal and vertical dotted lines represent
the mean metallicity and age, respectively.
\label{fig-4}}

\end{document}